\documentclass[11pt]{article}
\usepackage{cospar}
\usepackage{url}
\usepackage{graphicx}

\def\ergcms{\,erg cm$^{-2}$ s$^{-1}$}
\def\degr{\hbox{$^\circ$}}
\def\arcsec{\hbox{$^{\prime\prime}$}}

\title{A serendipitous survey for galaxy clusters by the
XMM-Newton Survey Science Center}

\author{A.D. Schwope\address{Astrophysikalisches Institut Potsdam
(AIP), An der Sternwarte 16, 14482 Potsdam, Germany},
        G. Lamer$^{1}$, 
        D. Burke\address{Harvard-Smithsonian Center for Astrophysics,
                60 Garden Street, Cambridge, MA 02138, USA},
        M. Elvis$^{2}$,
        M.G. Watson (on behalf of the XMM-SSC)\address{X-ray Astronomy
                Group, University of Leicester, University Road,
                Leicester LE1 7RH, UK},
        M.P. Schulze$^{1}$,
        G. Szokoly$^{1, }$\address{Max-Planck Institut f\"ur
        extraterrestrische Physik, 85748 Garching, Germany}
        and 
        T. Urrutia$^{1}$
}

\begin{document}

\maketitle

\begin{abstract}
We describe the initial results of a programme to detect and identify
extended X-ray sources found serendipitously in XMM-Newton observations. We
have analysed 186 EPIC-PN images at high galactic latitude with a
limiting flux of $1\times 10^{-14}$\,\ergcms\ and found 62 cluster
candidates. Thanks to the enhanced sensitivity of the XMM-Newton
telescopes, the new clusters found in this pilot study are on the average
fainter, more compact, and more distant than those found in previous
X-ray surveys. At our survey limit the surface density of
clusters is about 5\,deg$^{-2}$. We also present the first results of
an optical follow-up programme aiming at the redshift measurement of a
large sample of clusters. The results of this pilot study give a first
glimpse on the potential of serendipitous cluster science with
XMM-Newton based on real data. The largest, yet to be fulfilled
promise is the identification of a large number of high-redshift
clusters for cosmological studies up to $z=1$ or 1.5.
\end{abstract}

\section*{INTRODUCTION}

Clusters of galaxies are the largest gravitationally bound structures
in the universe and offer large diagnostic power to test cosmological
models. In current scenarios of structure formation, massive clusters
arise from gravitational instabilities in the extreme tail in the
distribution of density fluctuations. Hence, their number density and
their evolution depend critically on the initial distribution and the
cosmological parameters. The number density of clusters can be used to
constrain the amplitude of density fluctuations on small scales 
(quantified as $\sigma_8$), they may be used to estimate the ratio of
baryonic to non-baryonic matter in the universe from the observed
baryonic fraction in clusters, and finally, to constrain the matter density,
$\Omega_0$, by observing the number density evolution (White et
al.~1993, Gheller et al.~1998, Arnaud \& Evrard 1999, Wu \& Xue 2000,
Oukbir \& Blanchard 1992, Viana \& Liddle 1996, 1999, Henry 1997, Eke
et al.~1998, Borgani et al.~1999, Blanchard et al.~2000).

Therefore, much effort has been spent on the definition of
statistically complete samples of clusters and X-ray selection has
been proven to be one of the most efficient means to build such
samples. The main reasons for this are the high X-ray
luminosity that allows to detect them up to redshifts $z \simeq 1$ and beyond,
and the reduced ambiguity in the identification of a cluster in the
X-ray domain  
compared to the optical, where chance alignements play a much larger role. 
The first cluster samples were compiled already on the basis of UHURU,
Ariel V, and HEAO-1 data (Schwartz 1978, McHardy 1978, Piccinotti et
al.~1982, Kowalski et al.~1984) and comprised 76 clusters. Later on,
based on the Einstein Medium Sensitivity Survey with a survey area of
740\,deg$^2$, Henry et al.~(1992) established a catalog of clusters
with a flux limit two orders of magnitude deeper than the non-imaging
instruments could deliver with clusters reaching up to $z=0.58$. 
Clearly, most impacting was the ROSAT mission which revealed of order
1500 X-ray selected clusters of galaxies. The search for clusters
in the ROSAT data archive is still ongoing (see e.g.~Cruddace et
al.~2002, B\"ohringer et al.~2002, Ebeling et al.~2001, Rosati et al.~2002). 
An overview of the present knowledge of the luminosity-redshift plane
of clusters is given in Fig.~1. It compares three ROSAT-based cluster
surveys, the 
BCS (plus eBCS), the NORAS, and the CfA 160 deg$^2$ survey,
sorted by increasing survey sensitivity and/or decreasing sky
coverage (Ebeling et al. 1998, 2000;
B\"{o}hringer et al.~2000, Vikhlinin et al.~1998).
The ROSAT All-Sky Survey (RASS) with a survey area
of 20391 deg$^2$ and a flux limit of $\sim 2.4\times 10^{-12}$\ergcms\ 
yielded a surface density of clusters, $N < 0.03$ deg$^{-2}$.
The number of high-redshift clusters, $z > 0.4$, is low in
the RASS-based surveys and is also small in catalogues based on pointed
observations. Three quarter of the Viklinin et al.~clusters are below
redshift 0.4, the faintest clusters in their survey, at $1.6 \times 10^{-14}$,
is at about the limit of the pilot XMM-Newton based survey presented here.
The MACS survey (Ebeling et al.~2001), also based on the RASS,
explores the upper right corner of the $L_x
- z$ plane with considerable success. 
It is, however, not included in Fig.~1 since the catalogue is not
yet publicly available. 

\begin{figure}
\resizebox{75mm}{!}{\includegraphics[clip=]{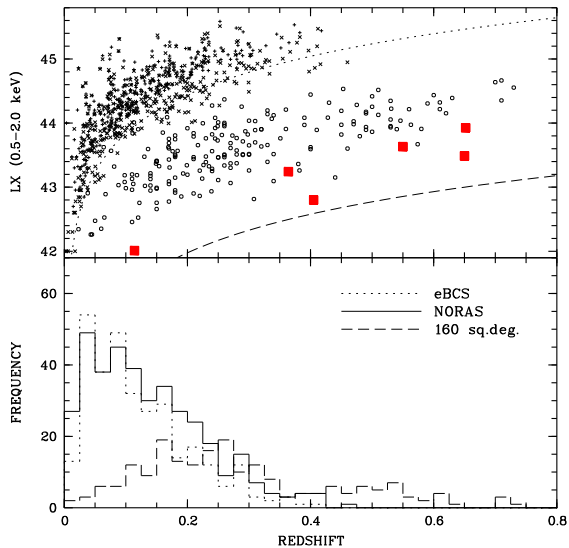}}
\hspace{5mm}
\parbox[b]{55mm}
{\caption{The upper panel shows the distribution of galaxy clusters in
the BCS (plus eBCs, $+$ symbols), the NORAS ($\times$), and the CfA 160
deg survey in the luminosity-redshift plane ($o$). 
The lower panel shows the corresponding redshift histograms. 
The solid line in the upper panel indicates the sensitivity
limit of the RASS. The 
dashed line gives a rough indication of the sensitivity limit of the
survey described here. Also shown in the upper panel are the loci of
six new XMM-clusters identified in the course of this programme.}}
\end{figure}

XMM-Newton has the highest throughput of any imaging X-ray observatory ever
flown. With its good spatial resolution, the unprecedented sensitivity of the
EPIC-cameras over a wide energy range (0.5 - 12 keV) and the large
field of view (30 arcmin) it detects about 50000 new X-ray sources per
year serendipitously. The Survey Science Center
(SSC) is a consortium of 10 European institutes with the
responsibility of creating the Software Analysis System SAS, compiling
the serendipitous catalogue of X-ray sources 
(the first version is due for release in early 2003,
http://xmmssc-www.star.le.ac.uk/newpages/xcat\_public.html) and
carrying out a follow-up and identification programme (XID). The whole
project is described by Watson et al.~(2001). In the XID-programme
the SSC is pursuing two closely related programmes: optical/near-IR
imaging of large numbers of XMM-Newton fields, and a `core' programme
to spectroscopically identify a significant sample of X-ray sources.
In the first phases of the programme we have concentrated mainly on
the identification of point-like X-ray sources (Barcons et al.~2002,
Watson et al.~2003), here we describe 
preliminary results of a study to characterize extended sources
in XMM-Newton images. 
The potential of the serendipitous source content of XMM-Newton for
cluster science was outlined by Romer et al.~(2001).

\section*{THE XMM-NEWTON SSC CLUSTER SURVEY}
The cluster project is embedded in the overall XID project
of the SSC, i.e.~the software and the energy bands for the source
detection are the same as for the pipeline processing of all data. 
The results will be made publicly available via the web-pages of the
SSC. In order to test the software and to optimize the strategy for
the production of a large sample of clusters, we have set up a pilot
study whose results are described here. 

\subsection*{Field selection}
For this pilot study we have selected observations for which the
observation PI granted SSC follow-up permission. This was declared
already on the original
proposal forms and applies to $\sim$75\% of all proposals. This
permission is granted to the non-target content of the observations,
hence, XMM-observations of known clusters are excluded from our pilot
study. We used images from the EPIC-PN camera only with the camera being
operated in the `FullFrame' or the `LargeWindow' mode. It is planned
later to combine the PN- and MOS-images, which will help to better
characterize the cluster morphology (bridging of CCD chip gaps) and
the spectral shape. We restricted ourselves to fields at 
high Galactic latitude, $|b| > 20\degr$, thus facilitating the
planned optical follow-up.
In order to gain in depth and to produce a tractable
number of possible fields we selected observations with nominal
exposure times $> 20$\,ksec. However, the time needed for the set up of the
PN-CCD and the loss of time due to bad space wheather lead to a
significant decrease of the usable exposure time. 
The application of the different selection criteria led to a total of
186 EPIC PN fields which were subsequently analysed. 
The fields are more or less homogeneously distributed in the sky and the 
actual observation time stretches from 2 ksec to 60 ksec with a peak
at 18 ksec (Fig.~2).

\begin{figure}
\resizebox{85mm}{!}{\includegraphics[width=85mm,clip=]{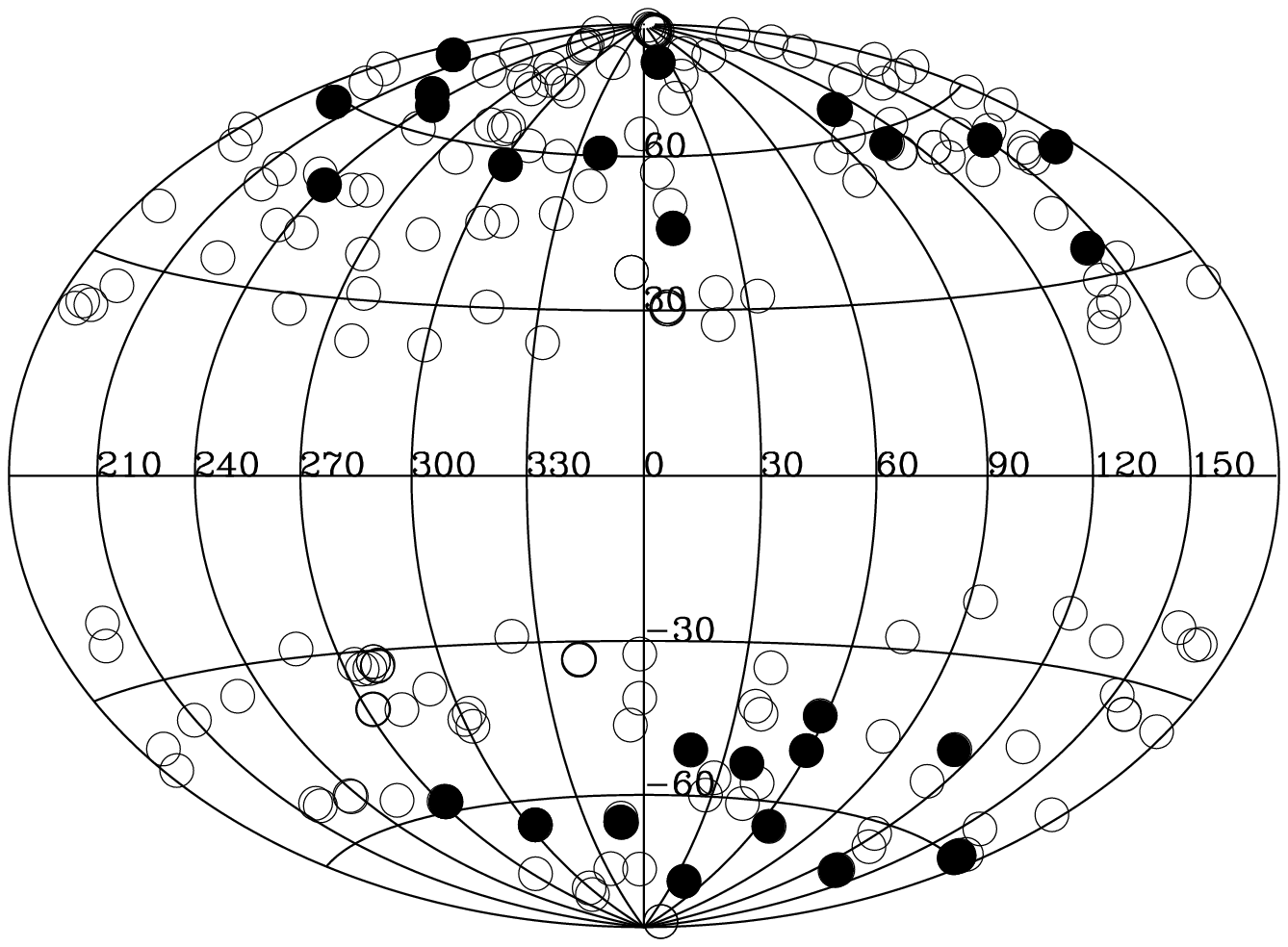}}
\hspace{5mm}
\resizebox{85mm}{!}{\includegraphics[width=85mm,clip=]{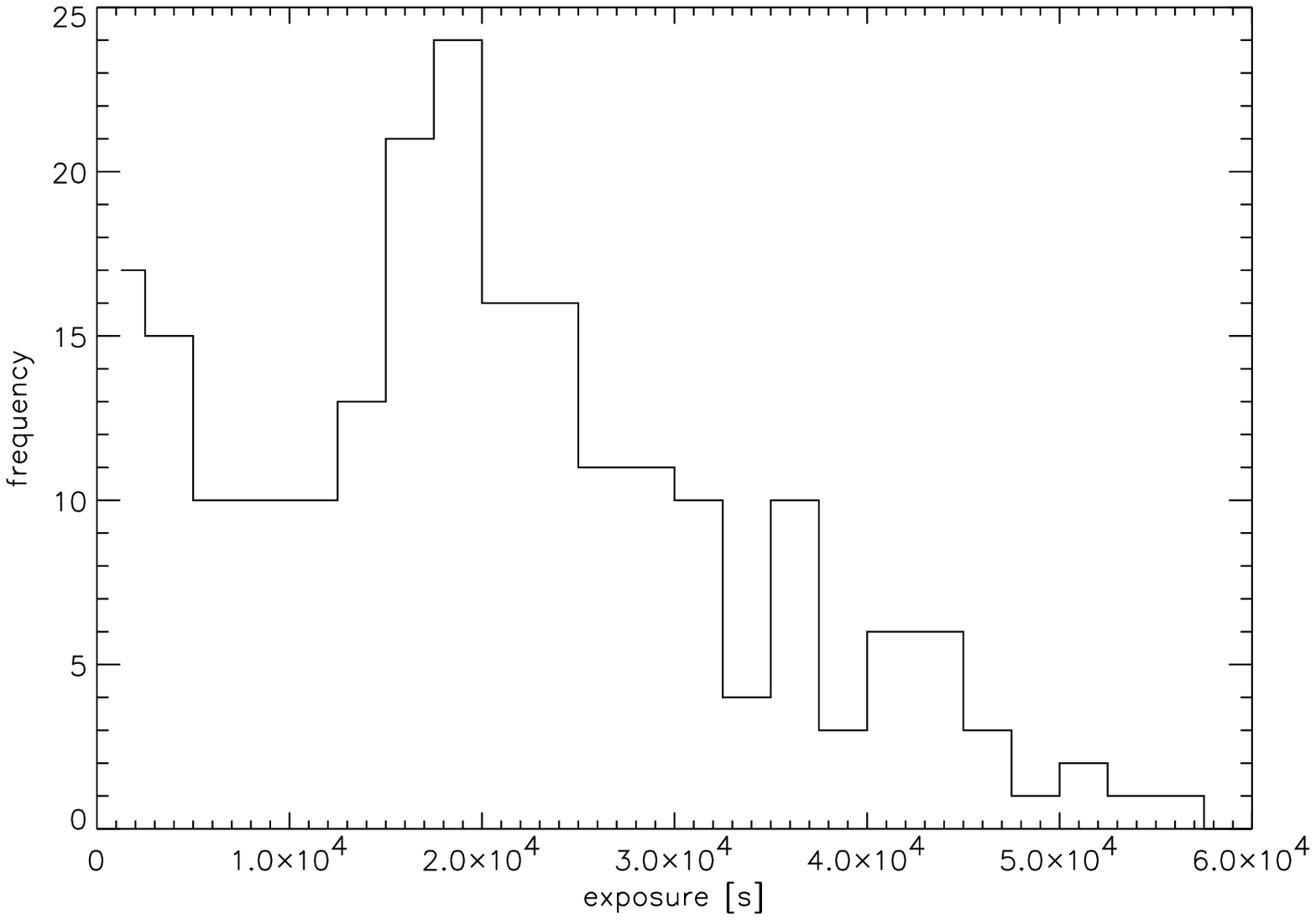}}
\resizebox{85mm}{!}{\includegraphics[width=85mm,clip=]{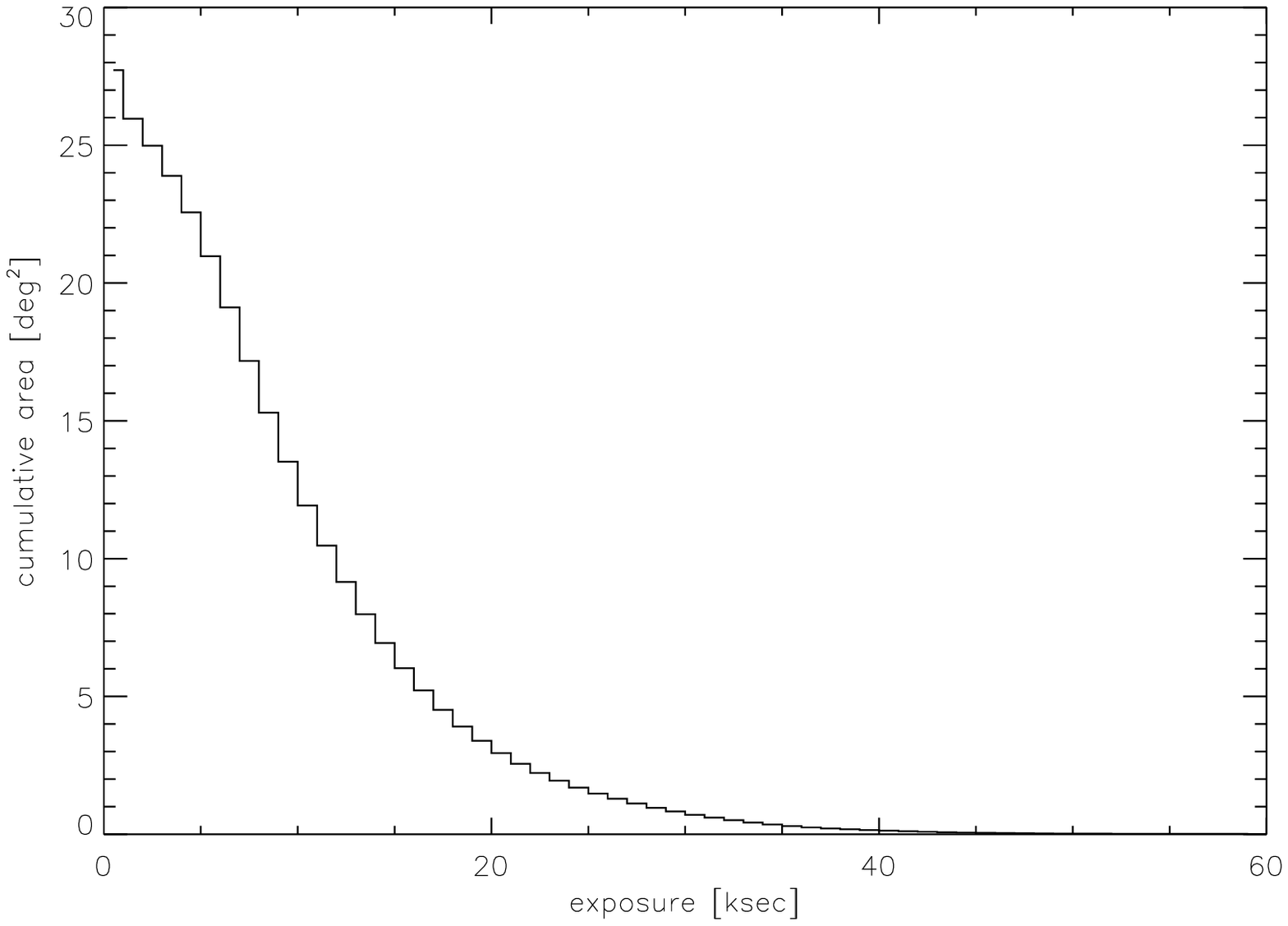}}
\hfill \parbox[b]{75mm}{
\caption{ ({\it top left}) Distribution of the 186 EPIC PN images in
the sky in galactic coordinates with newly detected clusters (extended X-ray sources)
indicated with filled symbols. 
({\it top right}) Distribution of the actual exposure time of the
PN-images.  ({\it bottom}) The vignetting-corrected survey area as a
function of exposure time.}}
\end{figure}

The geometric survey area of these observations is
$\sim 27.7$\,deg$^2$. The 
integrated vignetting corrected survey area of the 186 images,
taking into account only good time intervals is also shown in Fig.~2. 
Despite the large number of fields, the
survey area at 20 ksec true exposure, i.e.~at our nominal minimum
exposure time, is only 2.94 deg$^2$, roughly a factor of 10 below the
integrated geometric area.

\subsection*{Source search}
For each of the 186 observations, EPIC PN images were created in the
XID-band between 0.5--4.5 keV. 
The {\it edetect\_chain} of XMM SAS was used to search
for extended sources, and the task {\it emldetect} was 
used to fit extended model profiles to all detected sources. 
The source detection software combines a 
sliding-box detection and maximum-likelihood profile fitting with Gaussian
point-spread functions (PSF). The width of the PSF is adjusted 
according to the off-axis angle. 
The thresholds for detection likelihood and extent likelihood were set
to 10. The resulting fluxes of the extended sources were multiplied by
a factor of 1.25, which accounts for the
deviation of Gaussian-approximated cluster fluxes from the usually
better fitting King
profiles and was determined by detection runs on simulated PN images 
(Brunner \& Lamer 2003). Clearly, the derived X-ray fluxes are a first
approximation only, since the true cluster profile might also be different
from the assumed King profile (particularly
in dynamically young and irregular clusters), and since X-ray
point-sources are bound to be embedded in the diffuse cluster
emission. The fluxes of ROSAT clusters are typically given in the 0.5
-- 2.0 keV band. For the comparison between ROSAT and XMM-Newton clusters
(Figs.~1 and 3b), we applied a flux factor of 0.55, which accounts for the
different energy bands. This assumes a solar-abundance Raymond-Smith
spectrum of $kT_{\rm RS} = 4.8$\,keV, absorbed by cold interstellar
matter with $N_H = 4\times10^{20}$\,cm$^{-2}$.
The source lists were screened visually in order to remove
spurious detections. These originate mainly from imperfect background
determinations in the vicinity of CCD gaps or near bright or extended
X-ray sources.

\subsection*{Results}
\begin{figure}
\resizebox{85mm}{!}{\includegraphics{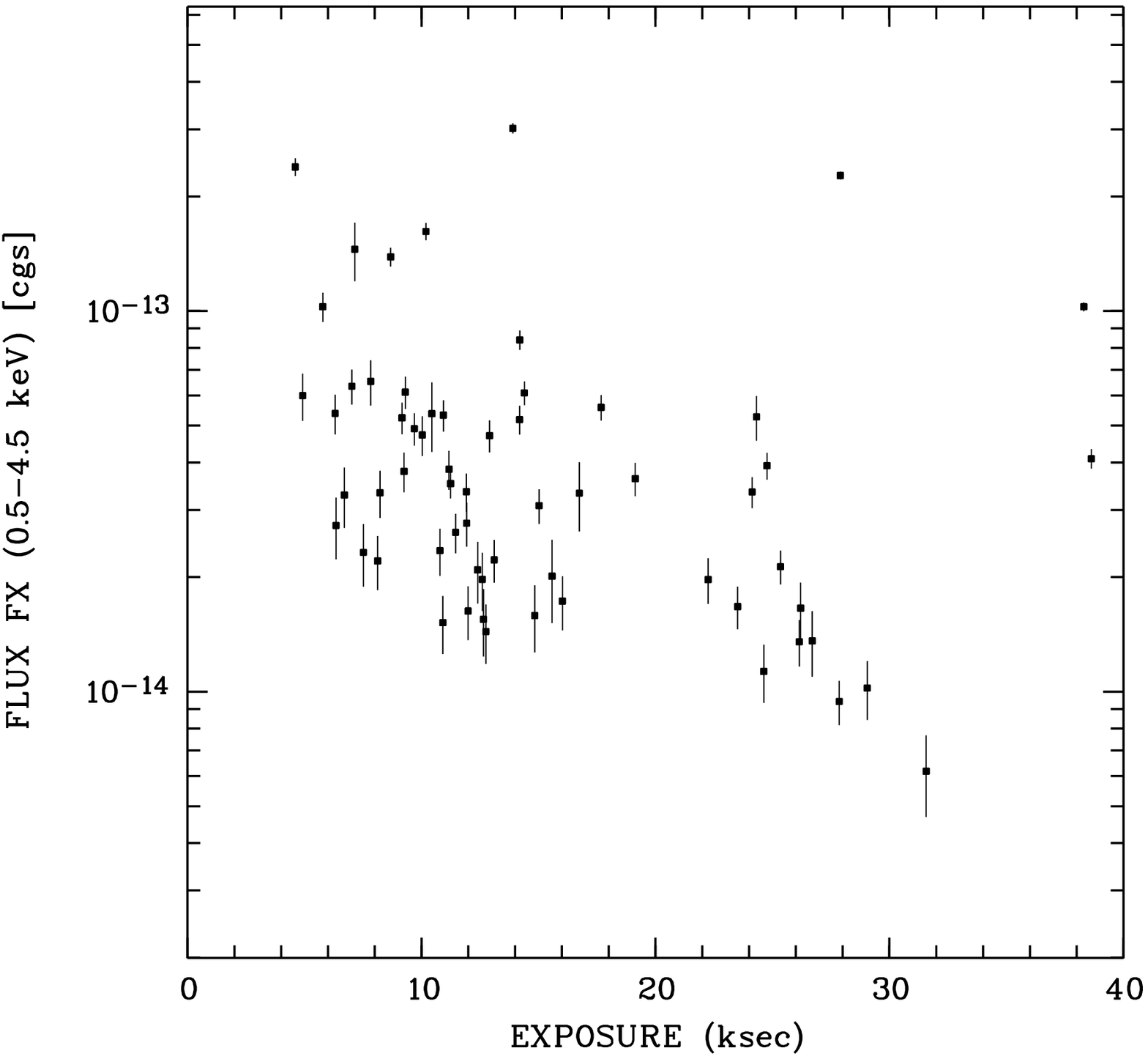}}
\hfill
\resizebox{85mm}{!}{\includegraphics[width=85mm,clip=]{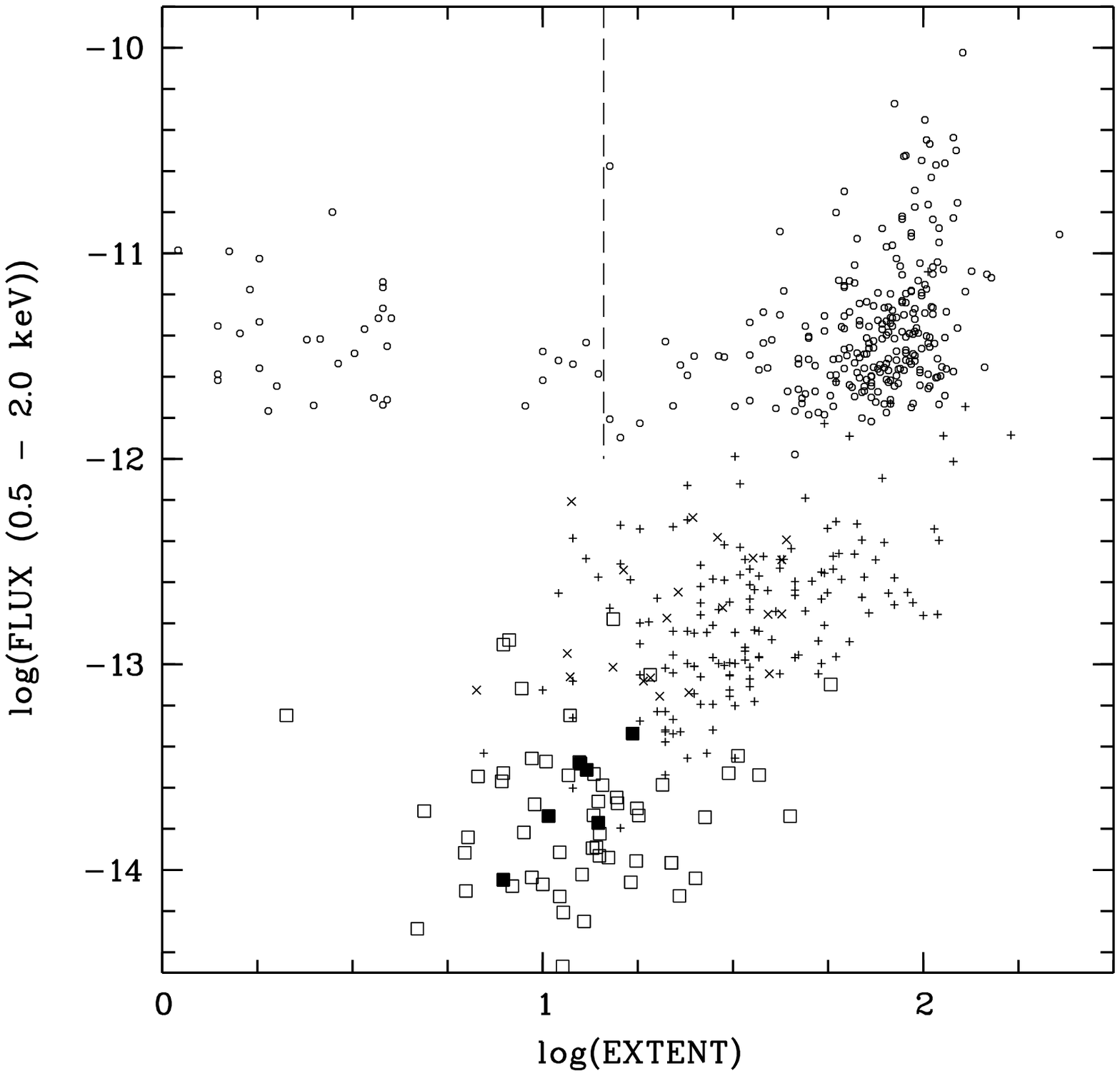}}
\caption{{\it (left)} X-ray flux (0.5--4.5 keV) of cluster candidates
as a function of vignetting-corrected exposure time.
{\it (right)} Distribution of newly found XMM-clusters in the 
X-ray-flux/extent plane (squares, filled symbols for clusters with
measured redshift) compared with clusters from the RBS
($\circ$), the WARPS ($\times$) and CfA 160 deg surveys ($+$). 
The quantity plotted which describes the extent of the XMM-clusters is
the Gaussian $\sigma$. 
For the ROSAT clusters the plotted quantities are: RBS -- the amount
by which the maximum likelihood fit is improved over a pointlike PSF
($\sim$14 arcsec, indicated by the vertical dashed line); WARPS \& CfA
160 deg -- the cluster core radius}
\label{label3}
\end{figure}

The limiting flux of the survey is 
$F_{\rm lim} \sim 10^{-14}$\,erg cm$^{-2}$ s$^{-1}$
in the $0.5-4.5$\,keV band, depending on exposure and extent of the 
sources. At this flux limit we find 62 clearly extended X-ray sources which
are regarded as cluster candidates. As expected, all new clusters are 
fainter than the RASS limit, the observed range is $7\times10^{-15} - 2.5
\times 10^{-13}$\,\ergcms. The median flux of the new clusters is $\sim
3\times 10^{-14}$\,\ergcms. About one third of the clusters were
detected in relatively shallow exposures below 10\,ksec, which means
that the planned full survey will make use of all suitable XMM-fields
disregarding the exposure time (see Fig.~\ref{label3}, left).
In the 3 deg$^{2}$ area with true exposure in excess of 20
ksec we find 15 clusters, which corresponds to a surface density of 5
deg$^{-2}$, more than a factor of 100 higher than at the limit of the
RASS. 

The median extent of the cluster
candidates is $\sim 13\arcsec$. In the present sample no correlation
between X-ray extent and X-ray flux is obvious. The right panel of
Fig.~\ref{label3} compares the X-ray fluxes and X-ray extent of the
new XMM-Newton clusters with different ROSAT-based X-ray surveys (RBS
- Schwope et al.~2000, CfA 160 deg -- Vikhlinin et al.~1998, WARPS --
Perlman et al.~2002). Although the quantities
plotted in the figures are not exactly the same for the different samples,
the comparison is nevertheless instructive.
The X-ray flux of the XMM-Newton clusters was transformed from the 
XID band (0.5 -- 4.5 keV) to the ROSAT band assuming a 4.8 keV
Raymond-Smith spectrum absorbed by a column with $N_H = 4\times
10^{20}$\,cm$^{-2}$. 
The X-ray extent of the XMM-Newton clusters is the Gaussian $\sigma$. 
The number which quantifies the extent of the RBS-clusters is the
extent parameter as found in the ROSAT Bright Source catalogue by
Voges et al.~(1999), i.e.~the Gaussian $\sigma$ which exceeds
the PSF of a pointlike source ($\sim$14 arcsec). RBS-clusters to the left
of the dashed line in the Fig.~3b are unresolved at the ROSAT-RASS resolution.
Since the ROSAT point-source PSF is small compared to the typical
extent of a RBS-cluster, the quantities plotted for the XMM- and the
RBS-clusters are nevertheless 
comparable (see the discussion of X-ray extent in the
RASS by Ebeling et al.~1998). The extent parameter plotted for the
WARPS and CfA 160 deg clusters is the cluster core radius as drawn
from the tables in Perlman et al.~(2002) and Viklinin et al.~(1998),
respectively. 

There is a pronounced trend from the bright, extended RBS-clusters
towards the faint, compact XMM-clusters.
The median extent of the new XMM-clusters is just 
compatible with the PSF of a point-like X-ray source in the RASS. 
Thanks to the good point-spread function and sensitivity of the
XMM-mirrors, faint, apparently compact X-ray selected galaxy clusters are
detected and can be studied. 
Presently we are exploring the completeness of our cluster detection
procedure. 
The {\it emldetect} task detects extended sources of size up
to 1/8 of a PN CCD, i.e.~up to 30\,arcsec, which limits 
the detection of extended, low-surface brightness structures. The {\it
  ewavelet} task will be used to complement source lists in future
versions of our catalogue.

\subsection*{Optical identifications}
Follow-up optical observations were initiated using facilities at ESO,
Calar Alto, the ING at La Palma and Las Campanas. Imaging data
were obtained in the framework of the optical imaging programme of the
SSC, mostly with the WFC at the 2.5m INT and with the WFI at the
ESO/MPG 2.2m telescope at La Silla (see
http://xmmssc-www.star.le.ac.uk/newpages/xid\_public.html). These
large-format, wide-field cameras provide photometry with limiting
magnitude of $V = 23- 24$\,mag and excellent astrometry. Follow-up
low-resolution spectroscopy of about 15 cluster candidates was performed
at the Magellan I telescope. Six of them belong to the sample
described in this paper. We
obtained long-slit spectra of the brightest 
cluster galaxies with the  LDSS-2 and the Boller \& Chivens spectrographs,
and determined redshifts in the range $z=0.07 - 0.65$. 
Fluxes and luminosities of the six new clusters in the present sample are
shown in comparison with ROSAT clusters in Figs.~1 and 3b. It is obvious
that the XMM-clusters extend the accessible parameter range in the 
$(L_X-z)$ and ($F_X - \sigma$) planes. 

\begin{figure}
\resizebox{59mm}{!}{\includegraphics[width=59mm,height=59mm,clip=]{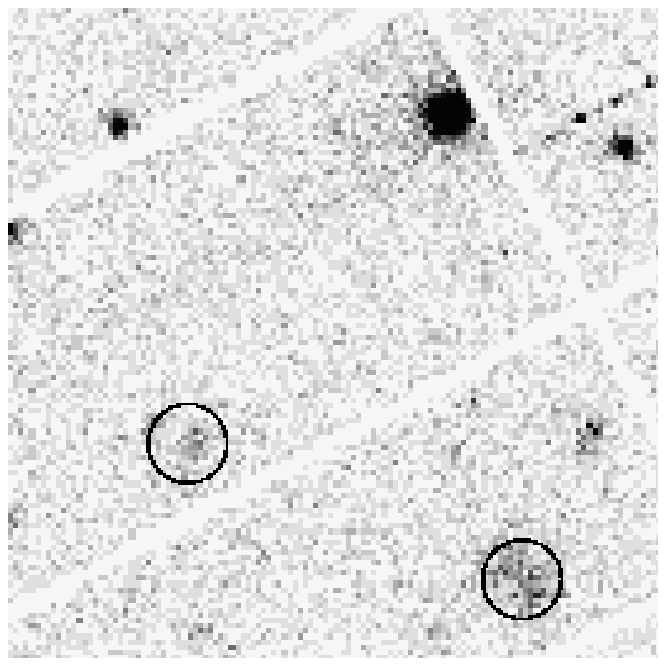}}
\hfill
\resizebox{59mm}{!}{\includegraphics[width=59mm,height=59mm,clip=]{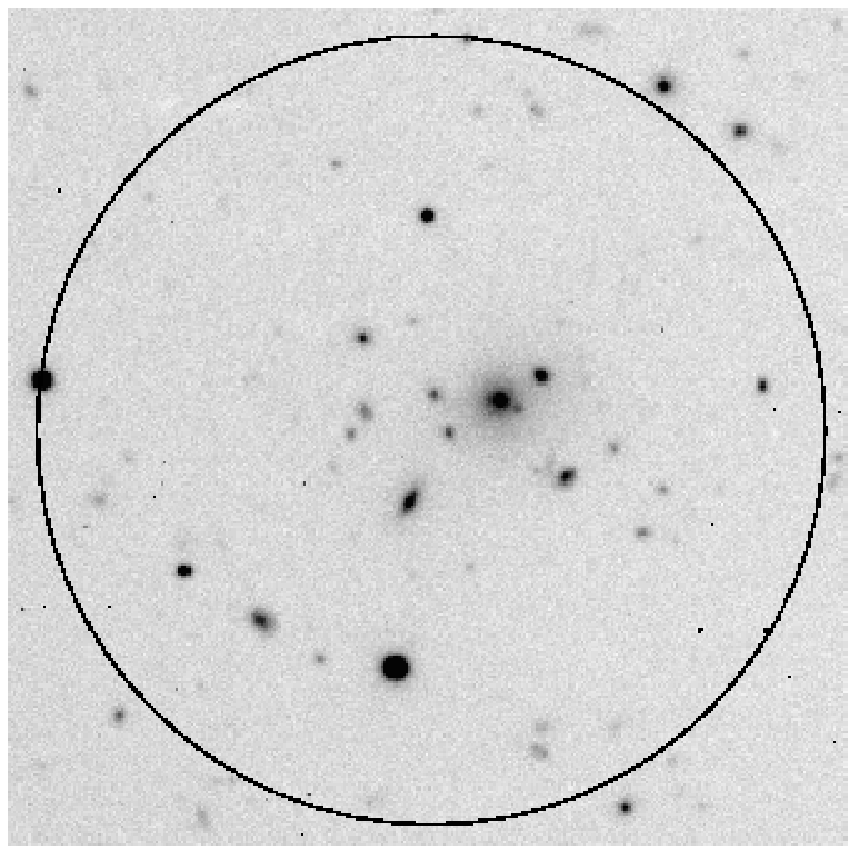}}
\hfill
\resizebox{59mm}{!}{\includegraphics[width=59mm,height=59mm,clip=]{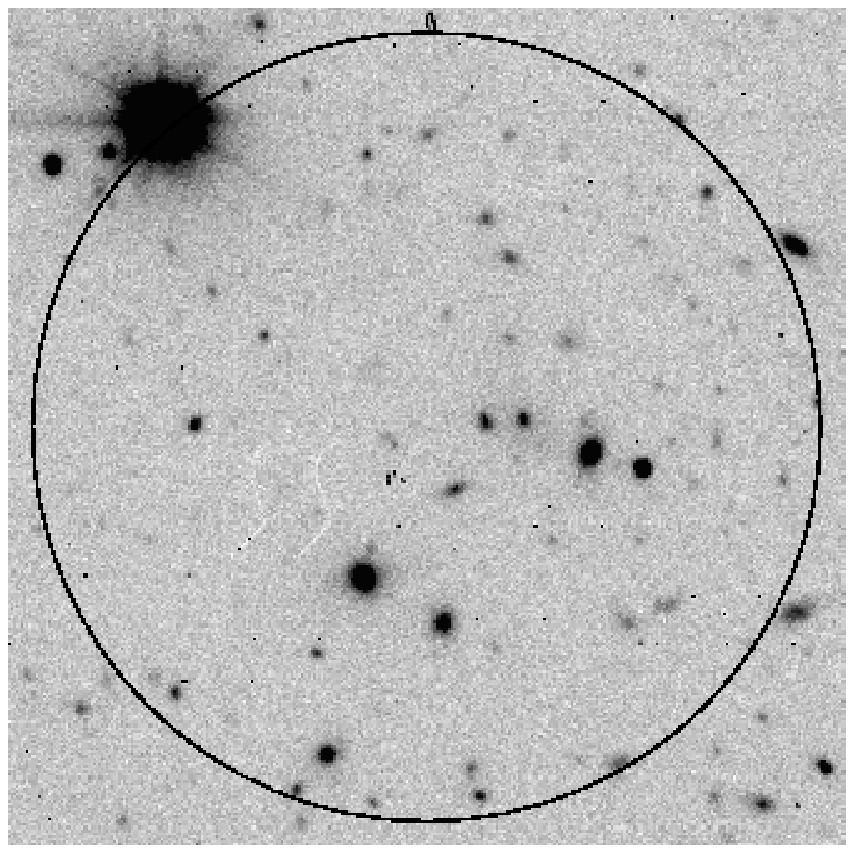}}
\caption{EPIC-PN (left) and VLT R-band images of the cluster pair discovered
in the field of WW Hor. The circles have a size of 30\,arcsec in all
three images. Spectra were taken of the two brightest cluster galaxies.
These were found at identical redshift of $z=0.65$.} \label{pair}
\end{figure}

Particularly intriguing is the discovery of paired clusters of
galaxies at high redshift, thus offering direct access to the study of
large-scale structures. In Fig.~\ref{pair} we show a cut-out of the
EPIC PN image of the field of WW Hor, observed in the CalPV phase
(revolution 181) for a total of 21148 sec with Full Frame through the
thin filter. The target, a cataclysmic variable star,  is the bright
point source in the upper part of the PN image. 
VLT R-band images of the two clusters are also
shown in the figure. Spectra with Magellan I were taken of the 
brightest galaxies in each cluster which gave identical redshifts of
$z=0.65$. At the distance of the two clusters, their angular
separation of 270\,arcsec corresponds to only a few diameters of the
clusters and is 
just $2 h_{50}^{-1}$\,Mpc. There are EPIC-PN fields with a much larger
number, $8-10$, of extended sources. Whether these
really form part of large-scale structures has to be tested by deep
multi-color imaging and dedicated spectroscopy.

\section*{SUMMARY AND OUTLOOK}
We have presented some initial results of a programme to understand
the extended source content in XMM-Newton EPIC PN images.
This project demonstrates the relative ease of creating large
samples of clusters of galaxies with XMM-Newton using X-ray extent 
as single selection criterion. The method will be particularly
valuable to find a significant number of clusters at high redshift and
to study cluster evolution up to redshifts of $z \sim 1 -
1.5$. We have shown already how the accessible parameter range in the
$F_X - \sigma$ plane is enlarged by the newly detected clusters and
we should be able to enlarge the parameter range
also in the $(L_X-z)$ plane.

Future work in the X-ray domain 
will concentrate on the optimisation of the
source detection, the inclusion of the MOS images, and a more
thorough determination of the survey area (taking into account the
lack of sensitivity for extended source detection in the vicinity of
bright and/or extended sources). Also more effort needs to be spent on
the determination of the source parameters, like temperature, extent,
and flux. The temperature of the X-ray emitting gas of the brighter
clusters will be determined by proper spectral fits and will be estimated
by X-ray hardness ratios for the clusters with low numbers of counts.  
A thorough optical
observation programme has been set up in order to determine
redshifts of the new clusters by photometry and spectroscopy.

\section*{ACKNOWLEDGEMENTS}
This work is based partly on observations with XMM-Newton, an ESA
Science Mission with instruments and contributions directly funded by
ESA member states and the USA (NASA).
This project is supported in part by
the German BMBF under DLR grant 50 OX 0201.
Based in part on observations performed with the ESO/WFI under
programmes 68.A-0473 and 69.A-0615.
We thank an anonymous referee for constructive criticism.

E-mail address of A.D. Schwope aschwope@aip.de
\end{document}